\documentclass[12pt]{article}
\makeatletter
\newcommand{\singlespacing}{\let\CS=\@currsize\renewcommand{\baselinestretch}{1}\tiny\CS}
\usepackage{epsfig}
\usepackage{graphicx}
\usepackage{verbatim}   % useful for program listings
\usepackage{color}      % use if color is used in text
\usepackage{amsfonts}
\usepackage{amsmath}
\usepackage{tikz}
\usepackage{float}
\usepackage{booktabs}
\usepackage{multirow}
\usepackage{caption}
\usepackage{subcaption}
\usepackage{array}% http://ctan.org/pkg/array
\usepackage{booktabs}
\usepackage{multirow}
\newcolumntype{d}{D{.}{.}{2.5}}
\newcolumntype{C}{>{\centering}p}
\begin{document}
\baselineskip=24pt
%\singlespacing
%\doublespacing
\parskip = 10pt
\def \qed {\hfill \vrule height7pt width 5pt depth 0pt}
\newcommand{\ve}[1]{\mbox{\boldmath$#1$}}
\newcommand{\IR}{\mbox{$I\!\!R$}}
\newcommand{\1}{\Rightarrow}
\newcommand{\bs}{\baselineskip}
\newcommand{\esp}{\end{sloppypar}}
\newcommand{\be}{\begin{equation}}
\newcommand{\ee}{\end{equation}}
\newcommand{\beanno}{\begin{eqnarray*}}
\newcommand{\inp}[2]{\left( {#1} ,\,{#2} \right)}
\newcommand{\eeanno}{\end{eqnarray*}}
\newcommand{\bea}{\begin{eqnarray}}
\newcommand{\eea}{\end{eqnarray}}
\newcommand{\ba}{\begin{array}}
\newcommand{\ea}{\end{array}}
\newcommand{\nno}{\nonumber}
\newcommand{\dou}{\partial}
\newcommand{\bc}{\begin{center}}
\newcommand{\ec}{\end{center}}
\newcommand{\2}{\subseteq}
\newcommand{\cl}{\centerline}
\newcommand{\ds}{\displaystyle}
\newcommand{\what}{\widehat}
\def\refhg{\hangindent=20pt\hangafter=1}
\def\refmark{\par\vskip 2.50mm\noindent\refhg}

\title{\sc A New Two Sample Type-II Progressive Censoring Scheme
 }
\author{\sc Shuvashree Mondal\footnote{Department of Mathematics and Statistics, Indian Institute of
Technology Kanpur, Pin 208016, India.}, Debasis Kundu \footnote{Department of Mathematics and Statistics, Indian Institute of
Technology Kanpur, Pin 208016, India.  Corresponding author.  E-mail: kundu@iitk.ac.in, Phone no. 91-512-2597141, Fax no. 91-512-2597500.
}}

\date{}
\maketitle
%\enlargethispage{1 in.}
\begin{abstract}
Progressive censoring scheme has received considerable attention in recent years.  In this paper we introduce a new type-II progressive 
 censoring scheme for two samples.  It is observed that the proposed censoring scheme is analytically more tractable than 
the existing joint progressive type-II censoring scheme proposed by Rasouli and Balakrishnan \cite{RB:2010}.  It has some other
advantages also.  We study the 
statistical inference of the unknown parameters based on the assumptions that the lifetime distribution of the experimental 
units for the two samples follow exponential distribution with different scale parameters.  The maximum likelihood estimators 
of the unknown parameters are obtained and their exact distributions are derived.  Based on the exact distributions of the maximum
likelihood estimators exact confidence intervals are also constructed.  For comparison purposes we have used bootstrap
confidence intervals also.  It is observed that the bootstrap confidence intervals work very well and they are very easy to implement in practice.  Some simulation experiments are performed to compare the performances of the proposed method with the 
existing one, and the performances of the proposed method are quite satisfactory.  One data analysis has been performed for illustrative 
purposes.  Finally we propose some open problems.

\end{abstract}



\newpage

\section{\sc Introduction}

Different censoring schemes are extensively used in practice to make a life testing experiment to be more time and cost effective.  In 
a type-I censoring scheme, the experiment is terminated at a prefixed time point.  But it may happen that, no failure is observed during that 
time and it will lead to a very poor statistical analysis of the associated model parameters.  To ensure a certain number of failures, type-II censoring scheme has been introduced in the literature.  But in none
of these censoring schemes any experimental unit can be removed  during the experiment.  The progressive censoring scheme
allows to withdraw some experimental units during the experiment also.  Different progressive censoring schemes have been introduced in the 
literature.  The most popular one is known as the progressive type-II censoring scheme and it can be briefly described as follows.
Suppose $n$ identical units are put on a life testing experiment.  The integer $k < n$ is prefixed, and $R_1$,\ldots,$R_k$ 
are $k$ prefixed non-negative integers such that $\ds \sum_{i=1}^{k}R_i +k=n$.  At the time of the first failure,  $R_1$ units are chosen randomly
from the remaining $n-1$ units and they are removed from the experiment.  Similarly at the time of the second failure, $R_2$ units 
are chosen randomly from the remaining $n-R_1-2$ units and they are removed, and so on.  Finally at the time of $k$-th failure 
remaining $R_k$ units are removed, and the experiment stops.  Extensive work has been done during the last ten years on various  
aspects of different progressive censoring schemes.  Interested readers may refer to the  recent book by Balakrishnan and Cramer 
\cite{BC:2014} for a detailed account on different progressive censoring schemes and the related issues.  See also 
Balakrishnan \cite{Bala:2007}, Pradhan and Kundu \cite{PK:2009} and Kundu \cite{Kundu:2008}, 
in this respect.

Although extensive work has been done on different aspects of the progressive censoring schemes for one sample, not much work has 
been done related to two sample problems.  Recently, Rasouli and Balakrishnan \cite{RB:2010} introduced the joint 
progressive type-II censoring for two samples.  The joint progressive censoring scheme is quite useful to compare the 
lifetime distribution of products from different units which are being manufactured by two different lines in the same facility.  
The joint progressive censoring (JPC) scheme introduced by Rasouli and Balakrishnan \cite{RB:2010} can be briefly 
stated as follows.  It is assumed that two samples of products of sizes $m$ and $n$, respectively, are selected from these two 
lines of operation (say Line 1 and Line 2), and they are placed on a life testing experiment simultaneously.  
A type-II progressive censoring scheme is implemented on the combined sample of size $N=m+n$ as follows.  Let $k < N$, and 
$R_1, \ldots, R_k$ are pre-fixed non-negative integers such that  $\ds \sum_{i=1}^k R_i + k = N$.  At the time of the first failure, 
it may be from Line 1 or Line 2, $R_1$ units are chosen at random from the remaining combined $N-1$ units which consists
of $S_1$ units from Line 1 and $T_1$ units from Line 2, and they are removed from the experiment.  Similarly at the the time of 
the second failure from the combined $N-2-R_1$ remaining units $R_2$ items are chosen at random, which consists of $S_2$ and $T_2$
units from Line 1 and Line 2, respectively, are removed, and so on.  Finally at the $k$-th failure remaining $R_k = S_k+T_k$ 
units are removed from the experiment, and the experiment stops.  Note that in a JPC, although $R_j$'s are pre-fixed, 
$S_j$'s and $T_j$'s are random quantities, and that makes the analysis more difficult.  Rasouli and Balakrishnan \cite{RB:2010}
provided the exact likelihood inference for two exponential populations under the proposed JPC scheme.    See also 
Parsi and Bairamov \cite{PB:2009}, Ashour and Abo-Kasem \cite{AA:2014}, Balakrishnan and Su 
\cite{BS:2015} for some problems related to the JPC scheme.

In this paper we introduce a new joint progressive type-II censoring (NJPC) scheme.  It
is observed that the proposed NJPC scheme is easier to handle analytically, therefore the properties of the proposed estimators
can be derived quite conveniently.  It has some other advantages also.  In this paper we provide the exact inference for 
two exponential populations under the NJPC scheme, although the results can be extended for other lifetime distributions 
also.  We obtain the maximum likelihood estimators (MLEs) of the unknown parameters
when it exist, and provide the exact distributions of the MLEs.  The generation of samples from the NJPC are quite simple, hence
the simulation experiments can be performed quite conveniently.  It is observed that the MLEs obtained from the NJPC scheme 
satisfy the stochastic monotonicity properties stated by Balakrishnan and Iliopoulos \cite{BI:2009}, hence the exact distribution 
of the MLEs can be used
to construct the confidence intervals of the unknown parameters.  For comparison purposes we proposed to use bootstrap confidence
intervals also.  Some simulation experiments are performed to compare the performances of the estimators based on JPC and NJPC.  It is
observed that the estimators based on NJPC behave better than the corresponding estimators based on JPC for certain censoring 
schemes.  One data analysis has been performed for illustrative purposes.  

The rest of the paper is organized as follows.  In Section 2  we introduce the model and provide the necessary assumptions.  
The MLEs are obtained and their exact distributions are provided in Section 3.  In Section 4 we provide a simple algorithm to 
simulate data from a NJPC scheme and obtain the expected time of the experiment.  The construction of confidence intervals are provided in Section 5.  Simulation results and 
the analysis of one data set are provided in Section 6.  Finally in Section 7 we propose some open problems and conclude the 
paper.

\section{\sc Model Description and Model Assumption}

Suppose we have products from two different populations.  We draw a random sample of size $m$ from population one (Pop-1) 
and a random sample of size $n$ from population two (Pop-2).  We place two independent samples simultaneously on a life 
testing experiment.  The proposed NJPC can be described as follows.  Let $k < \min\{m, n\}$ be the total number of failures 
to be observed and $R_1, \ldots, R_{k-1}$ are such that $\ds \sum_{i=1}^{k-1}(R_i+1) < \min \{m, n \}$.  
Suppose the first failure takes place at the time point $W_1$ and it comes from Pop-1, then $R_1$ units are randomly chosen 
from the remaining $m-1$ surviving units of Pop-1 and they are removed.  At the same time $(R_1 +1)$ units are randomly chosen 
from $n$ surviving units of Pop-2 and they are removed.  Suppose the next failure takes place at the time point    
$W_2$ and it comes from Pop-2, then $R_2+1$ units are chosen at random from the remaining $m-1-R_1$ surviving units of Pop-1, 
and they are removed.  At the same time $R_2$ units are chosen at random from the remaining $n-2-R_1$ surviving units of Pop-2, 
and they are removed, and so on.  Finally, at the time of the $k$-th failure, it may be either from Pop-1 or from Pop-2, 
all the remaining items from both the populations 
are removed and the experiment stops.  

We further define a new set of random variables $Z_1, \ldots, Z_k$, where $Z_j$ = 1 
if the $j$-th failure takes place from Pop-1 and $Z_j$ = 0, otherwise.  Hence for a NJPC scheme, the data will be of the form 
$(\bold W,\bold Z)$, where $W=(W_1, \ldots, W_k)$, $W_1\leq \ldots \leq W_k$ and $Z=(Z_1, \ldots, Z_k)$.  Schematically, NJPC
can be described as follows.

\noindent Case-I: $k$-th failure comes from Pop-1

\vspace{5mm}

\begin{tikzpicture}[scale=0.8]
\draw[gray, thick] (-7,4)node[anchor=north]{\textbf{Pop-1}} -- (2,4);
%\draw[gray,thick](2,4)--(3,5)--(2.7,3.3)--(3.5,4)--(10,4);
\draw[gray, thick](2,4)--(2.5,3.5)--(2.8,4.5)--(3.1,4)--(10,4);

\draw[gray, thick] (-7,0)node[anchor=north]{\textbf{Pop-2}} -- (2,0);
\draw[gray, thick](2,0)--(2.5,-.5)--(2.8,.5)--(3.1,0)--(10,0);

\filldraw[black] (-7,4) circle (2pt) node[anchor= south] {\small start};
\filldraw[black] (-7,0) circle (2pt) node[anchor= south] {\small start};

\draw[dashed,->] (-5,0) --(-4,2) node[anchor=west] {\small $R_1+1$};
\draw[arrows=->] (-5,4)--(-4,6)node[anchor=west] {\small $R_1$};

\filldraw[black](-4,1.5) circle(0.0005pt) node[anchor=west]{\small withdrawn};
\filldraw[black](-4,5.5) circle(0.0005pt) node[anchor=west]{\small withdrawn};
\draw [dashed] (-5,-1)--(-5,5);

\filldraw[black] (-5,4) circle (2pt) node[anchor= north west] {$W_1$};

\draw[arrows=->] (-1,0) --(0,2) node[anchor=west]{\small $R_2$};
\filldraw[black] (-1,0) circle(2pt) node[anchor=north west]{$W_2$};
\draw[dashed,->] (-1,4)--(0,6) node[anchor=west]{\small $R_2+1$};

\filldraw[black](0,1.5) circle(0.0005pt) node[anchor=west]{\small withdrawn};
\filldraw[black](0,5.5) circle(0.0005pt) node[anchor=west]{\small withdrawn};

\draw [dashed] (-1,-1)--(-1,5);

\draw[dashed,->] (5,0) --(6,2) node[anchor=west]{\small $n-\sum_{j=1}^{k-1}(R_j+1)$};
e\draw[arrows=->] (5,4)--(6,6) node[anchor=west]{\small $m-\sum_{j=1}^{k-1}(R_j+1)-1$};
\draw [dashed] (5,-1)--(5,5);
\filldraw[black] (5,4) circle (2pt) node[anchor= north west] {$W_k$};
\filldraw[black](6,1.5) circle(0.0005pt) node[anchor=west]{\small withdrawn};
\filldraw[black](6,5.5) circle(0.0005pt) node[anchor=west]{\small withdrawn};

\end{tikzpicture}

\newpage
\noindent Case-II: $k$-th failure comes from Pop-2

\vspace{5 mm}

\begin{tikzpicture}[scale=0.8]
\draw[gray, thick] (-7,4)node[anchor=north]{\textbf{Pop-1}} -- (2,4);
%\draw[gray,thick](2,4)--(3,5)--(2.7,3.3)--(3.5,4)--(10,4);
\draw[gray, thick](2,4)--(2.5,3.5)--(2.8,4.5)--(3.1,4)--(10,4);

\draw[gray, thick] (-7,0)node[anchor=north]{\textbf{Pop-2}} -- (2,0);
\draw[gray, thick](2,0)--(2.5,-.5)--(2.8,.5)--(3.1,0)--(10,0);

\filldraw[black] (-7,4) circle (2pt) node[anchor= south] {\small start};
\filldraw[black] (-7,0) circle (2pt) node[anchor= south] {\small start};

\draw[dashed,->] (-5,0) --(-4,2) node[anchor=west] {\small $R_1+1$ };
\draw[arrows=->] (-5,4)--(-4,6)node[anchor=west] {\small $R_1$};
\draw [dashed] (-5,-1)--(-5,5);
\filldraw[black] (-5,4) circle (2pt) node[anchor= north west] {$W_1$};
\filldraw[black](-4,1.5) circle(0.0005pt) node[anchor=west]{\small withdrawn};
\filldraw[black](-4,5.5) circle(0.0005pt) node[anchor=west]{\small withdrawn};

\draw[arrows=->] (-1,0) --(0,2) node[anchor=west]{\small $R_2$};

\filldraw[black] (-1,0) circle(2pt) node[anchor=north west]{$W_2$};

\draw[dashed,->] (-1,4)--(0,6) node[anchor=west]{\small $R_2+1$};
\draw [dashed] (-1,-1)--(-1,5);
\filldraw[black](0,1.5) circle(0.0005pt) node[anchor=west]{\small withdrawn};
\filldraw[black](0,5.5) circle(0.0005pt) node[anchor=west]{\small withdrawn};

\draw[arrows=->] (5,0) --(6,2) node[anchor=west]{\small $n-\sum_{j=1}^{k-1}(R_j+1)-1$};
\filldraw[black] (5,0) circle (2pt) node[anchor= north west] {$W_k$};

\draw[dashed,->] (5,4)--(6,6) node[anchor=west]{\small $m-\sum_{j=1}^{k-1}(R_j+1)$};
\draw [dashed] (5,-1)--(5,5);
\filldraw[black](6,1.5) circle(0.0005pt) node[anchor=west]{\small withdrawn};
\filldraw[black](6,5.5) circle(0.0005pt) node[anchor=west]{\small withdrawn};

\end{tikzpicture}

Suppose $X_1, \ldots, X_m$ denote the lifetimes of $m$ units of Pop-1, and it is assumed that they are independent and identically 
distributed (i.i.d.) exponential random variables with mean $\theta_1$ (Exp($\theta_1$)).  Similarly, it is assumed that  
$Y_1,\ldots ,Y_n$ denote the lifetimes of $n$ units of Pop-2, and they are i.i.d exponential random variables with mean 
$\theta_2$.

\section{\sc Maximum likelihood estimators And Their Exact Distributions}

\subsection{\sc Maximum Likelihood Estimators }

For a given sampling scheme $m$, $n$, $k$ and $\ds R_1, \ldots, R_{k-1}$ based on the observation $(\bold W,\bold Z)$ 
the likelihood function can be written as
\begin{eqnarray}
L(\theta_1,\theta_2| \bold w,\bold z)=C\frac{1}{\theta_1^{m_k}}\frac{1}{\theta_2^{n_k}} e^{-(\frac{A_1}{\theta_1}+\frac{A_2}{\theta_2})}; 
\label{ll}
\end{eqnarray}
where the normalizing constant $\ds C=\prod_{i=1}^{k}[(m-\sum_{j=1}^{i-1}(R_j+1))z_i+(n-\sum_{j=1}^{i-1}(R_j+1))(1-z_i)]$, 
$\ds A_1=\sum_{i=1}^{k-1}(R_i+1)w_i+(m-\sum_{i=1}^{k-1}(R_i+1))w_k$, $\ds A_2=\sum_{i=1}^{k-1}(R_i+1)w_i+(n-\sum_{i=1}^{k-1}(R_i+1))w_k$, 
$\ds m_k=\sum_{i=1}^{k}z_i$, $n_k=\sum_{i=1}^{k}(1-z_i) = k - m_k$.  From \eqref{ll} it follows that $(m_k, n_k, A_1, A_2)$ is the 
joint complete sufficient statistics of the unknown parameters $(\theta_1, \theta_1)$.  It is immediate that the MLEs of both 
$\theta_1$ and $\theta_2$ exist when $1\leq m_k \leq k-1$, and they are as follows:
$$
\widehat{\theta}_1 = \frac{A_1}{m_k} \ \ \ \ \hbox{and} \ \ \ \ \widehat{\theta}_2 = \frac{A_2}{n_k}.
$$
Hence $(\widehat{\theta}_1, \widehat{\theta}_2)$ is  the conditional MLE of $(\theta_1,\theta_2$),
conditioning on $1\leq m_k \leq k-1$.

\subsection{\sc Joint and Marginal Distributions}

In this section we provide the joint and marginal distribution function of $\widehat{\theta}_1$ and $\widehat{\theta}_2$ 
based on the joint and marginal moment generating function (MGF) approach.  Lemma 1 is needed for further development.

\noindent {\sc Lemma 1:} 
$$
P(m_k=r)=\sum_ {\substack{\bold z}\in Q_r}\{{\prod_{i=1}^{k}\frac{(m-\sum_{j=1}^{i-1}(R_j+1))z_i+(n-\sum_{j=1}^{i-1}(R_j+1))(1-z_i)}{(m-\sum_{j=1}^{i-1}(R_j+1))\theta_2+(n-\sum_{j=1}^{i-1}(R_j+1))\theta_1}}\}\theta_1^{k-r}\theta_2^r,
$$
where $\ds Q_r=\left \{\bold z=(z_1,\ldots,z_k):\sum_{i=1}^{k}z_i=r \right \}; r=0,\ldots,k.$

\noindent {\sc Proof:} See in the Appendix.   \qed

Note that when $m = n$, then
\be
P(m_k=r) = {k\choose{r}} \left ( \frac{\theta_2}{\theta_1+\theta_2} \right )^r 
\left ( \frac{\theta_2}{\theta_1+\theta_2} \right )^{k-r}; \ \ \ r = 0, \ldots, k.    \label{pmk}
\ee

Now we provide the joint moment generating function (MGF) of $(\widehat{\theta}_1, \widehat{\theta}_2)$ conditioning on 
$1 \le m_k \le k-1$.

\noindent {\sc Theorem 1:} The joint MGF of $(\widehat{\theta}_1, \widehat{\theta}_2)$ conditioning on 
$1 \le m_k \le k-1$ is given by  
\begin{equation}
M_{ \what{\theta}_1, \what{\theta}_2 }\big(t_1,t_2\big) = \frac{\sum_{r=1}^{k-1}P(m_k=r)
\prod_{s=1}^{k}({1-\alpha_{sr} t_1-\beta_{sr} t_2})^{-1}}{P(1\leq m_k \leq k-1)},
\end{equation}
where 
\beanno
\alpha_{sr} & = & 
\frac{(m-\sum_{i=1}^{s-1}(R_i+1))\theta_1\theta_2}{r\{(m-\sum_{i=1}^{s-1}(R_i+1))\theta_2+(n-\sum_{i=1}^{s-1}(R_i+1))\theta_1\}}   \\
\beta_{sr} & = & 
\frac{(n-\sum_{i=1}^{s-1}(R_i+1))\theta_1\theta_2}{(k-r)\{(m-\sum_{i=1}^{s-1}(R_i+1))\theta_2+(n-\sum_{i=1}^{s-1}(R_i+1))\theta_1\}}.
\eeanno
\noindent {\sc Proof:} See in the Appendix.   \qed

\noindent Using Theorem 1, we immediately get the following corollary.

\noindent {\sc Corollary 1:} Conditioning on $1\leq m_k \leq k-1$, the marginal MGF of $\widehat{\theta}_1$ and 
$\widehat{\theta}_2$ are given by 
\beanno
M_{\what{\theta}_1}\big(t\big) & = & \frac{\sum_{r=1}^{k-1}P(m_k=r)
\prod_{s=1}^{k}({1-\alpha_{sr} t})^{-1}}{P(1\leq m_k \leq k-1)} \ \ \ \hbox{and} \ \ \  \\
M_{\what{\theta}_2}\big(t\big) & = & \frac{\sum_{r=1}^{k-1}P(m_k=r)
\prod_{s=1}^{k}({1-\beta_{sr} t})^{-1}}{P(1\leq m_k \leq k-1)},
\eeanno
respectively.

Hence we have the PDFs of $\widehat{\theta}_1$ and $\widehat{\theta}_2$ as follows.

\noindent {\sc Theorem 2:} Conditioning on $1\leq m_k \leq k-1$, the PDF of   
$\what{\theta}_1$ is given by 
\begin{equation}
f_{\what{\theta}_1}\big(t\big)=\frac{\sum_{r=1}^{k-1}P(m_k=r)
g_{X_r}(t)}{P(1\leq m_k \leq k-1)}.
\end{equation}
Here $\ds X_r \overset{d}{=} \sum_{s=1}^{k} U_{sr}$, where
$U_{sr} \sim \hbox{Exp}(\alpha_{sr})$ and they are independently distributed.  Also, $g_{X_r}(t)$ is the PDF of $X_r$, and
when $m \ne n$,
$$
g_{X_r}(t) = \prod_{s=1}^{k}\frac{1}{\alpha_{sr}} \times \sum_{s=1}^{k}\frac{e^{-\frac{t}{\alpha_{sr}}}}{\prod_{j=1,j\neq s}^{k}
(\frac{1}{\alpha_{jr}}-\frac{1}{\alpha_{sr}})}; \ \ \  t > 0,
$$
and 0, otherwise.  When $m = n$, 
$$
g_{X_r}(t) = \frac{1}{\Gamma (k) \alpha_r^k} t^{k-1} e^{-\frac{t}{\alpha_r}}; \ \ \ t > 0,
$$
and 0, otherwise.  Here $\ds \alpha_r = \frac{\theta_1 \theta_2}{r(\theta_1+\theta_2)}$.  

The PDF of $\widehat{\theta}_2$ 
is given by 
\begin{equation}
f_{\what{\theta}_2}\big(t\big)=\frac{\sum_{r=1}^{k-1}P(m_k=r)
g_{Y_r}(t)}{P(1\leq m_k \leq k-1)}.
\end{equation}
Here $Y_r\overset{d}{=}\sum_{s=1}^{k}V_{sr}$, where
$V_{sr} \sim \hbox{Exp}(\beta_{sr})$ and they are independently distributed.  Also, $g_{Y_r}(t)$ is the PDF of $Y_r$, and 
when $m \ne n$,
$$
g_{Y_r}(t) = \prod_{s=1}^{k}\frac{1}{\beta_{sr}} \times \sum_{s=1}^{k}\frac{e^{-\frac{t}{\beta_{sr}}}}{\prod_{j=1,j\neq s}^{k}(\frac{1}
{\beta_{jr}}-\frac{1}{\beta_{sr}})}; \ \ \ t > 0,
$$
and 0, otherwise.  When $m = n$,
$$
g_{Y_r}(t) = \frac{1}{\Gamma (k) \beta_r^k} t^{k-1} e^{-\frac{t}{\beta_r}}; \ \ \ t > 0,
$$
and 0, otherwise.  Here $\ds \beta_r = \frac{\theta_1 \theta_2}{(k-r)(\theta_1+\theta_2)}$.

\noindent {\sc Proof:} It immediately follows from Corollary 1.    \qed
  
\noindent {\sc Remark:} The distribution of the MLE is a mixture of $k-1$ components, where each component is  
a sum of $k$ independent exponentially distributed random variables.   When $m = n$, it is a weighted mixture of
gamma distributions.

We can easily obtain the moments of $\widehat{\theta}_1$ and $\widehat{\theta}_2$.  When $m \ne n$, the first two moments are
\begin{eqnarray*}
E(\widehat{\theta}_1) & = & \frac{\sum_{r=1}^{k-1}P(m_k=r) \sum_{s=1}^{k} \alpha_{sr}}{P(1\leq m_k \leq k-1)}   \\
E({\widehat{\theta}_1}^2) & = & \frac{\sum_{r=1}^{k-1}P(m_k=r)
(2\sum_{s=1}^{k} \alpha_{sr}^2+\sum_{\substack{i\neq j}} \alpha_{ir} \alpha_{jr})}{P(1\leq m_k \leq k-1)}     \\
E(\widehat{\theta}_2) & = & \frac{\sum_{r=1}^{k-1}P(m_k=r)
\sum_{s=1}^{k}\beta_{sr}}{P(1\leq m_k \leq k-1)}     \\
E({\widehat{\theta}_2}^2) & = & \frac{\sum_{r=1}^{k-1}P(m_k=r)
(2\sum_{s=1}^{k}\beta_{sr}^2+\sum_{\substack{i \neq j}} \beta_{ir} \beta_{jr})}{P(1\leq m_k \leq k-1)}.  
\end{eqnarray*}
When $m = n$,
$$
E(\widehat{\theta}_1)=\frac{\sum_{r=1}^{k-1}P(m_k=r)k\alpha_r}{P(1\leq m_k \leq k-1)} \ \ \ \hbox{and} \ \ \
E({\widehat{\theta}_1}^2)=\frac{\sum_{r=1}^{k-1}P(m_k=r)k(k+1){{\alpha_r}^2}}{P(1\leq
m_k \leq k-1)}
$$
$$
E(\widehat{\theta}_2)=\frac{\sum_{r=1}^{k-1}P(m_k=r)k\beta_r}{P(1\leq m_k \leq k-1)}  \ \ \ \hbox{and} \ \ \
E({\what{\theta}_2}^2)=\frac{\sum_{r=1}^{k-1}P(m_k=r)k(k+1){{\beta_r}^2}}{P(1\leq
m_k \leq k-1)}.
$$
Here $\alpha_r$ and $\beta_r$ are same as defined before, and $P(m_k=r)$ is given by (\ref{pmk}).

\begin{figure}[H]
\begin{subfigure}{.5\textwidth}
  %\centering
  \includegraphics[width=.8\linewidth]{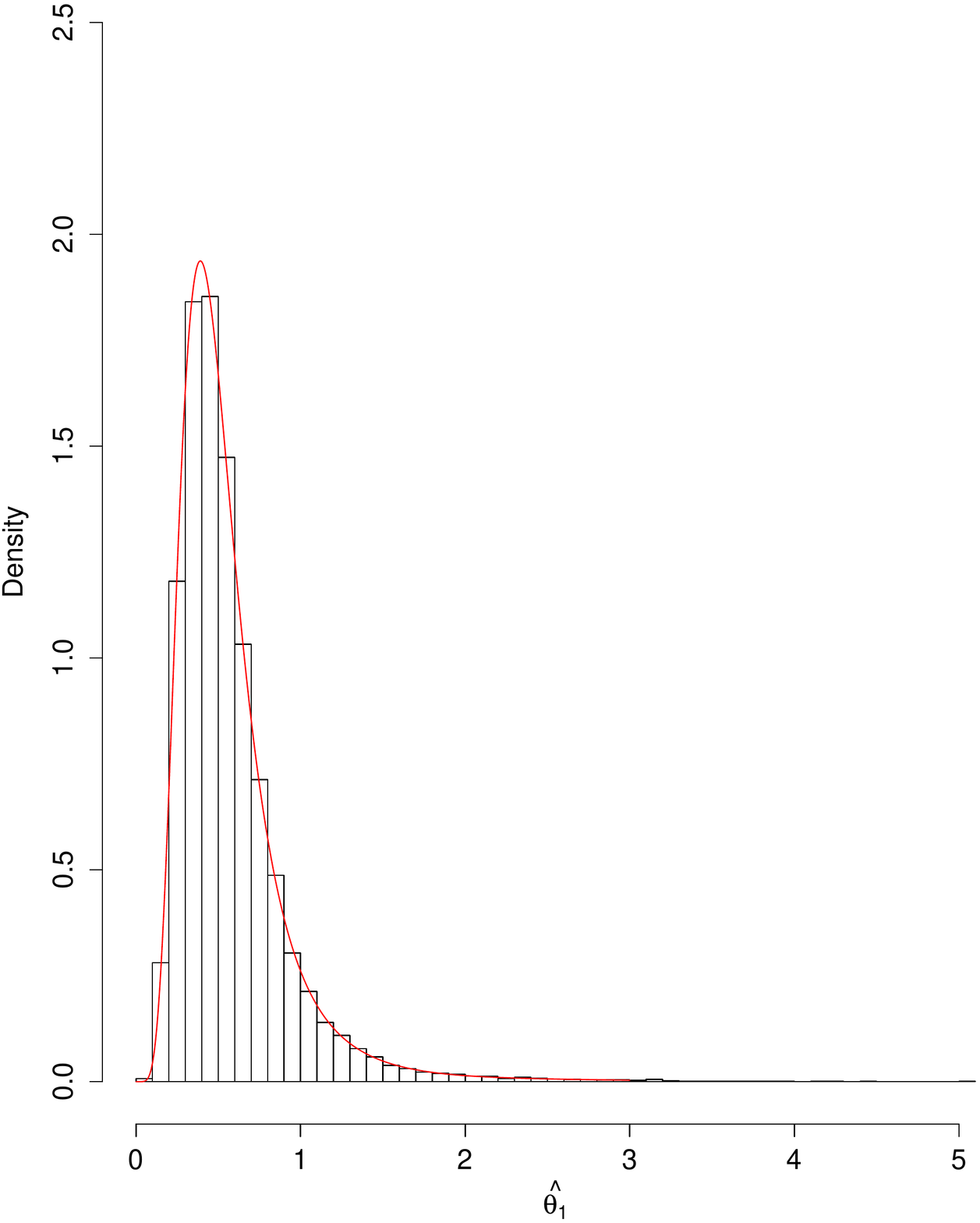}
  \caption{Histogram of $\what{\theta}_1$ along with its PDF}
  \label{fig:sfig1}
\end{subfigure}%
\begin{subfigure}{.5\textwidth}
  %\centering
  \includegraphics[width=.8\linewidth]{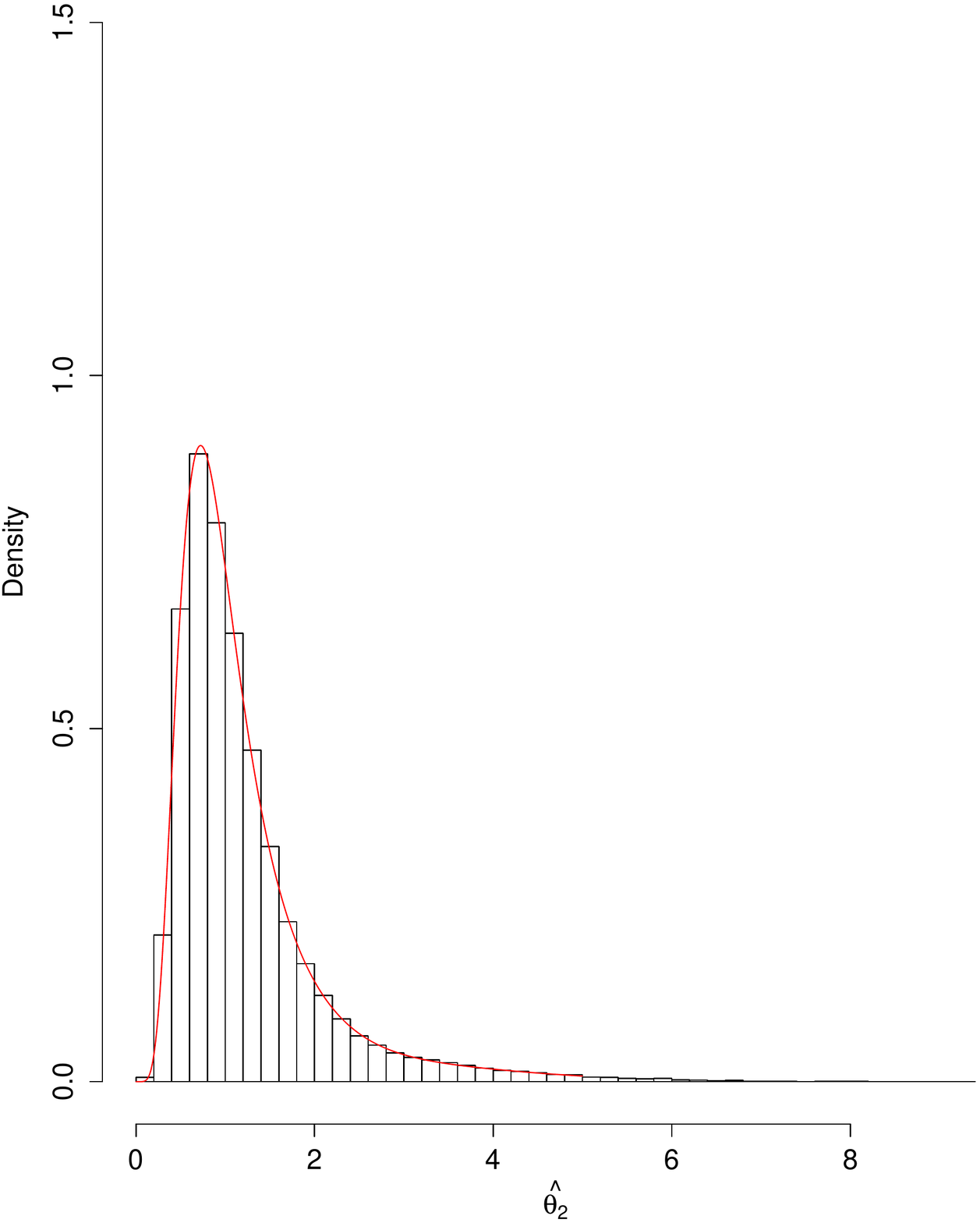}
  \caption{Histogram of $\what{\theta}_2$ along with its PDF}
  \label{fig:sfig2}
\end{subfigure}
\caption{Histogram of $\what{\theta}_1$ and $\what{\theta}_2$ along with its PDF,taking $\theta_1=.5$, $\theta_2=1$, $m=20$, $n=25$, $k=8$,$R=(7,0_{(6)})$}
\label{fig:fig1}
\end{figure}

\begin{figure}[H]
\begin{subfigure}{.5\textwidth}
  %\centering
  \includegraphics[width=.8\linewidth]{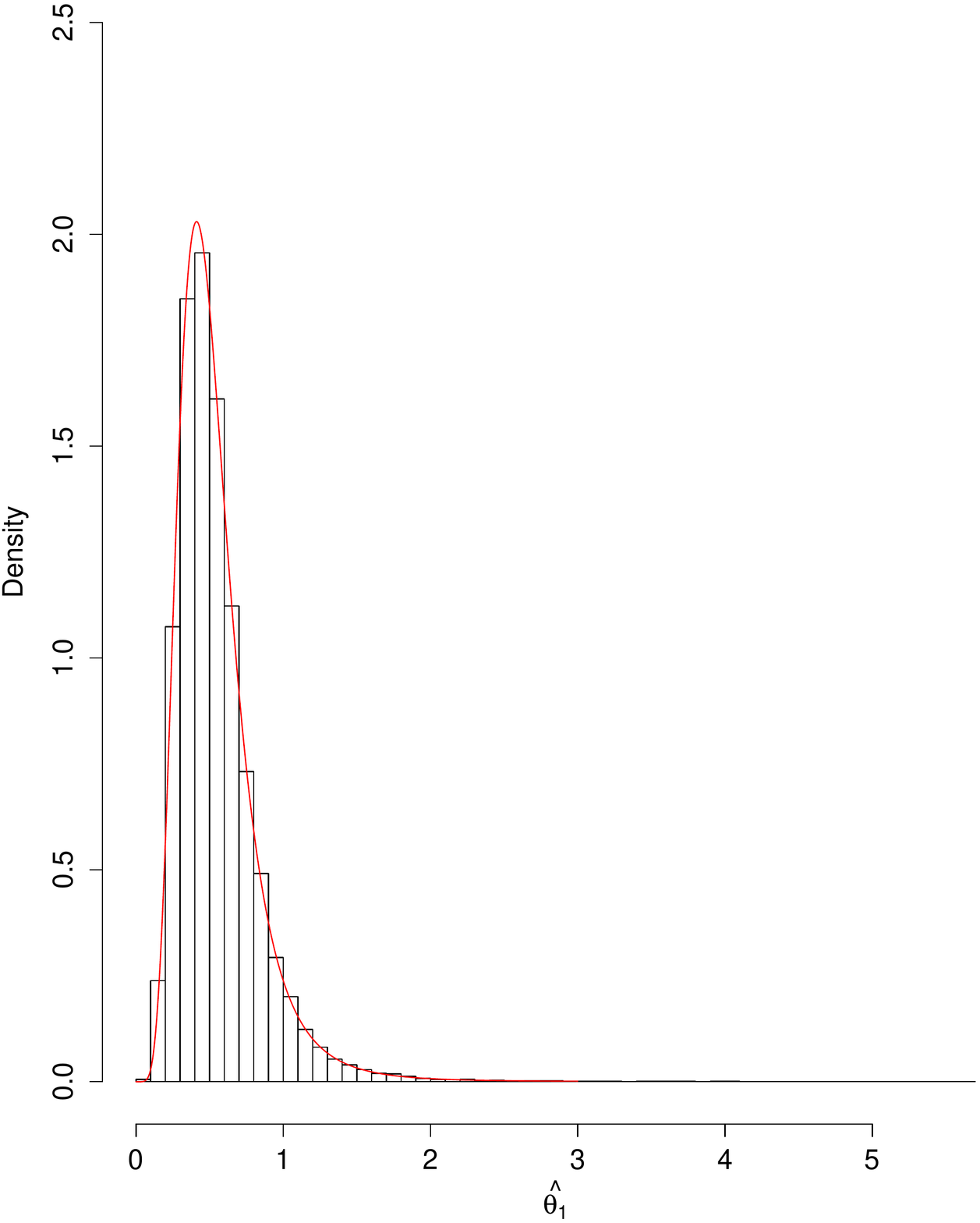}
  \caption{Histogram of $\what{\theta}_1$ along with its PDF}
  \label{fig:sfig1}
\end{subfigure}%
\begin{subfigure}{.5\textwidth}
  %\centering
  \includegraphics[width=.8\linewidth]{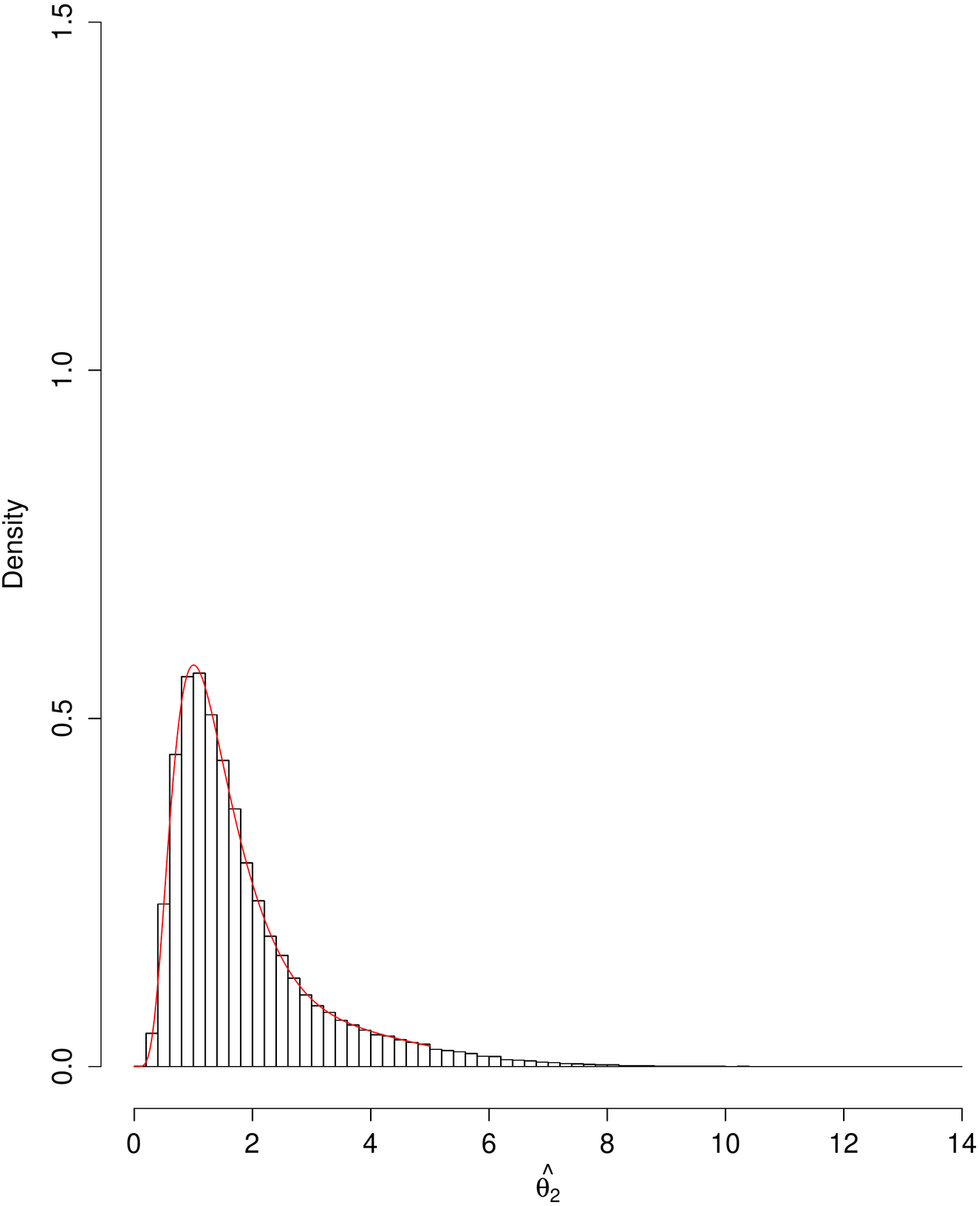}
  \caption{Histogram of $\what{\theta}_2$ along with its PDF}
  \label{fig:sfig2}
\end{subfigure}
\caption{Histogram of $\what{\theta}_1$ and $\what{\theta}_2$ along with its PDF,taking $\theta_1=.5$, $\theta_2=1.5$, $m=20$, $n=25$, $k=8$,$R=(7,0_{(6)})$}
\label{fig:fig2}
\end{figure}

\begin{figure}[H]
\begin{subfigure}{.5\textwidth}
  %\centering
  \includegraphics[width=.8\linewidth]{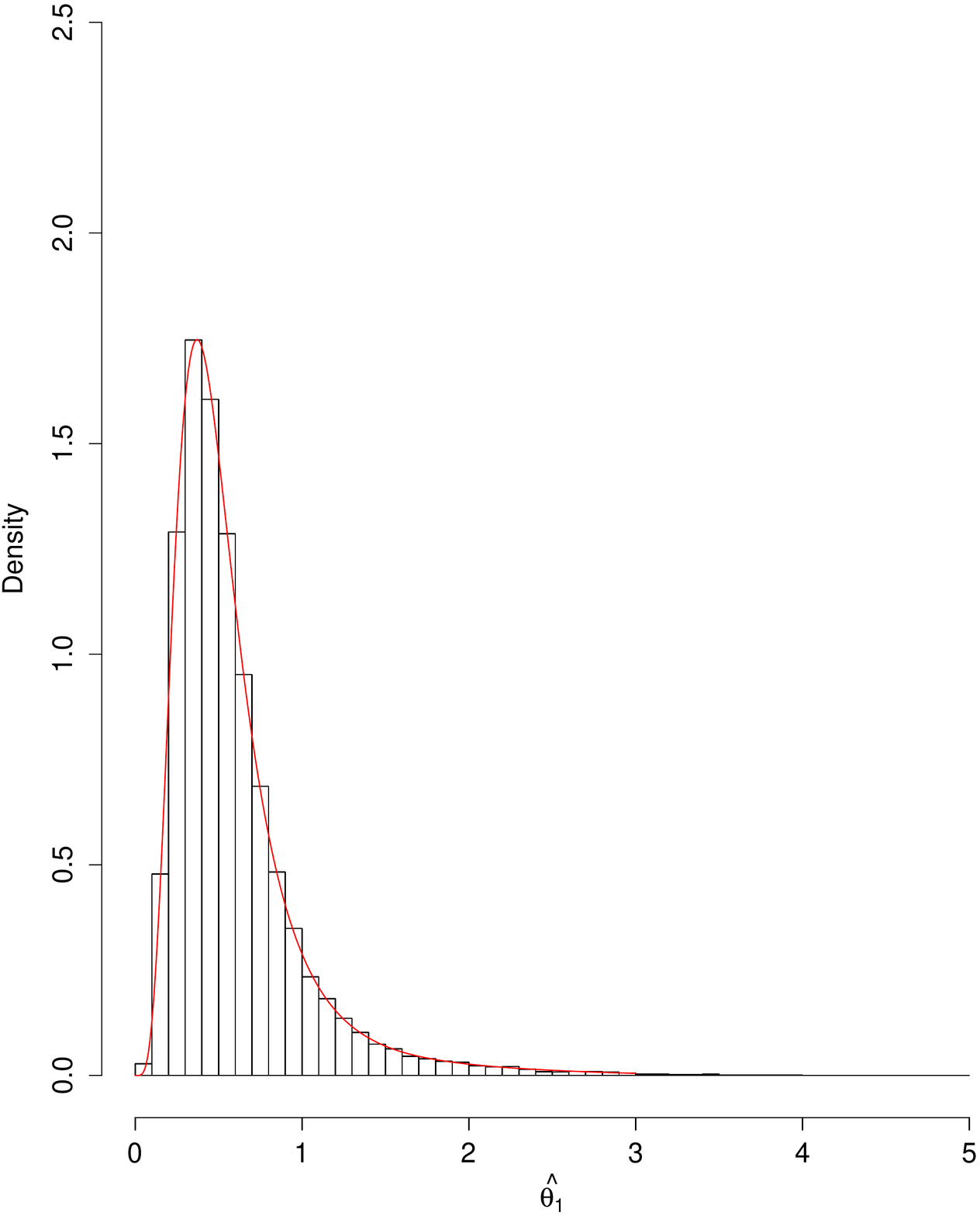}
  \caption{Histogram of $\what{\theta}_1$ along with its PDF}
  \label{fig:sfig1}
\end{subfigure}%
\begin{subfigure}{.5\textwidth}
  %\centering
  \includegraphics[width=.8\linewidth]{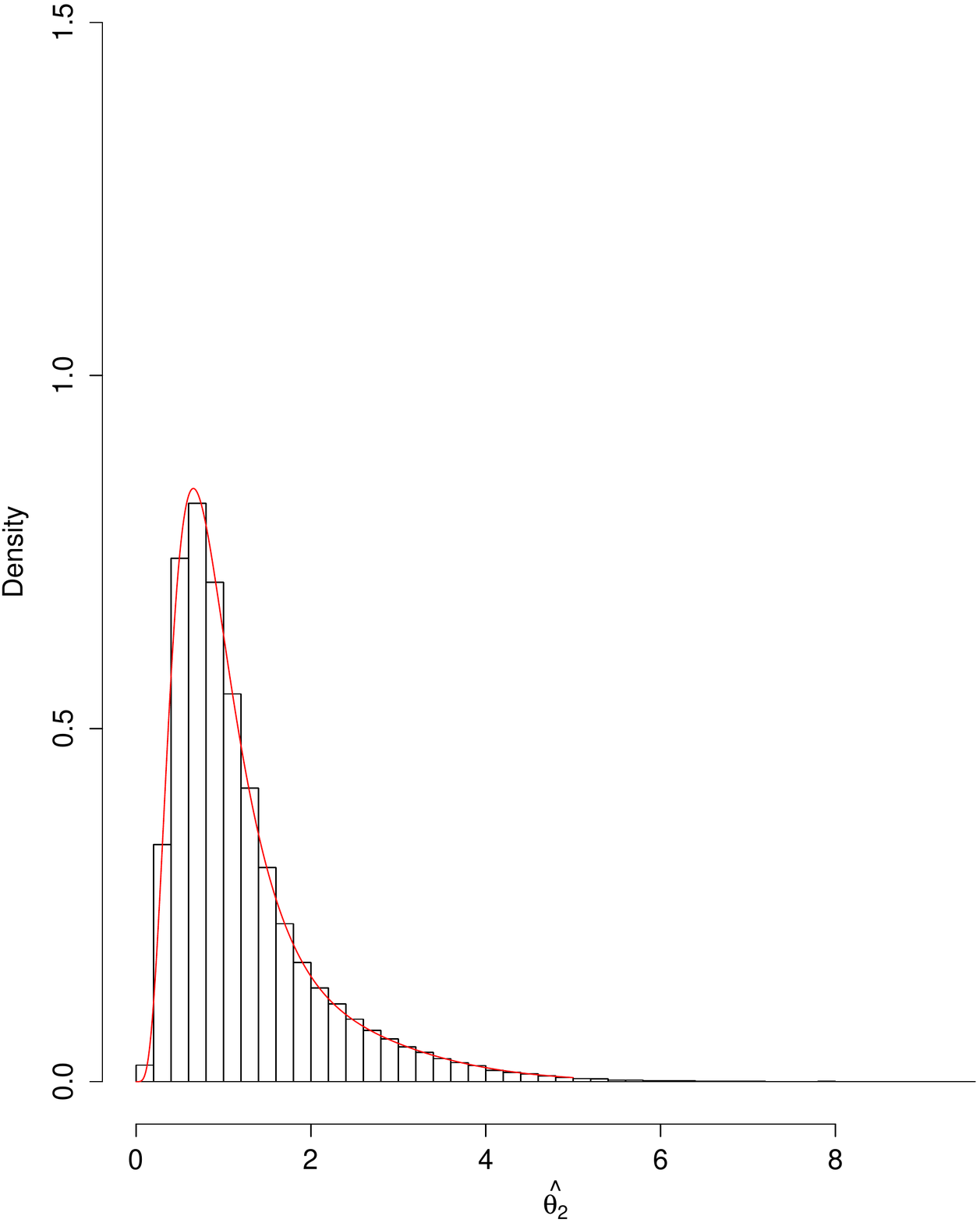}
  \caption{Histogram of $\what{\theta}_2$ along with its PDF}
  \label{fig:sfig2}
\end{subfigure}
\caption{Histogram of $\what{\theta}_1$ and $\what{\theta}_2$ along with its PDF,taking $\theta_1=.5$, $\theta_2=1$, $m=20$, $n=25$, $k=6$, $R=(2_{(5)})$}
\label{fig:fig3}
\end{figure}

\begin{figure}[H]
\begin{subfigure}{.5\textwidth}
  %\centering
  \includegraphics[width=.8\linewidth]{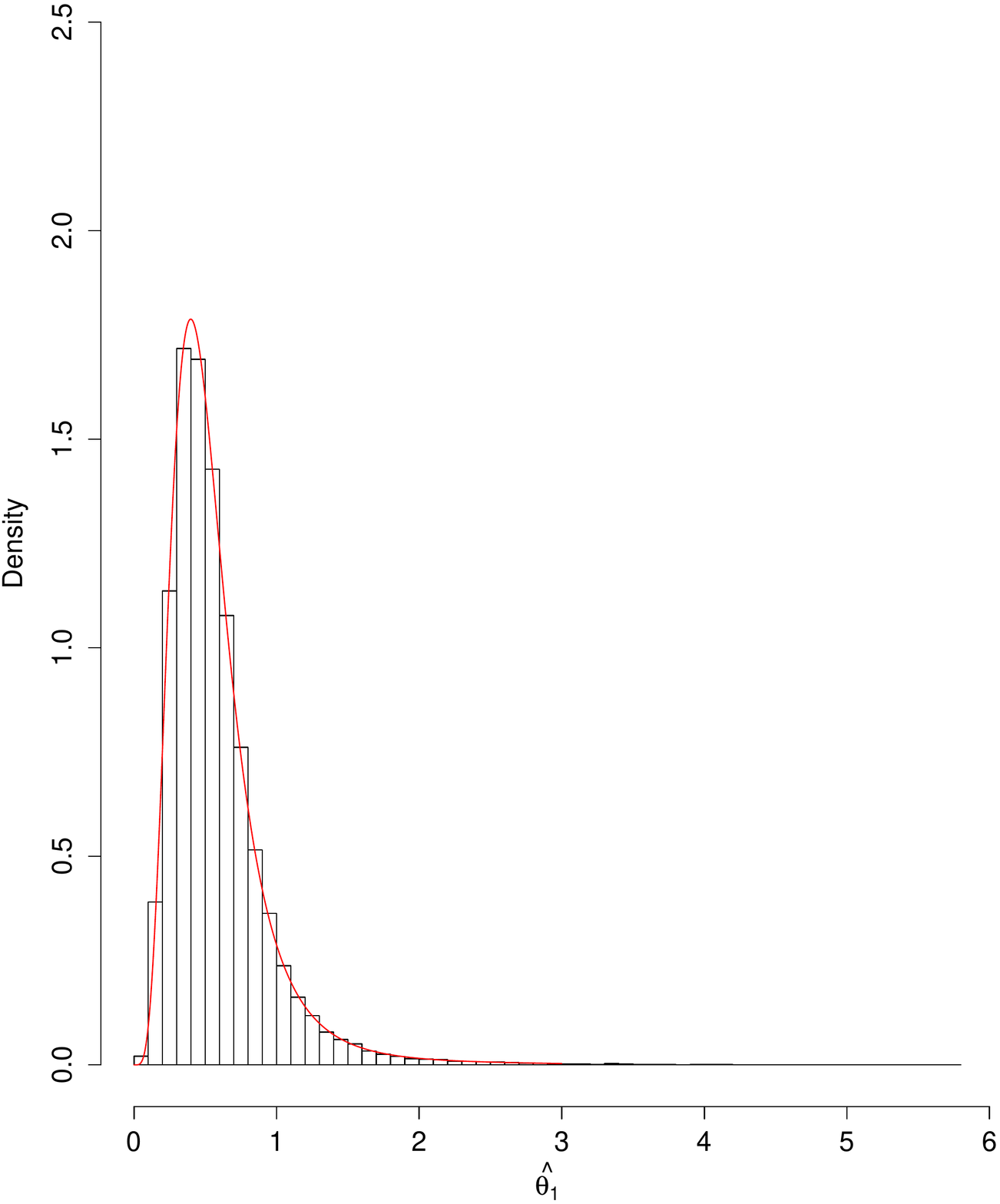}
  \caption{Histogram of $\what{\theta}_1$ along with its PDF}
  \label{fig:sfig1}
\end{subfigure}%
\begin{subfigure}{.5\textwidth}
  %\centering
  \includegraphics[width=.8\linewidth]{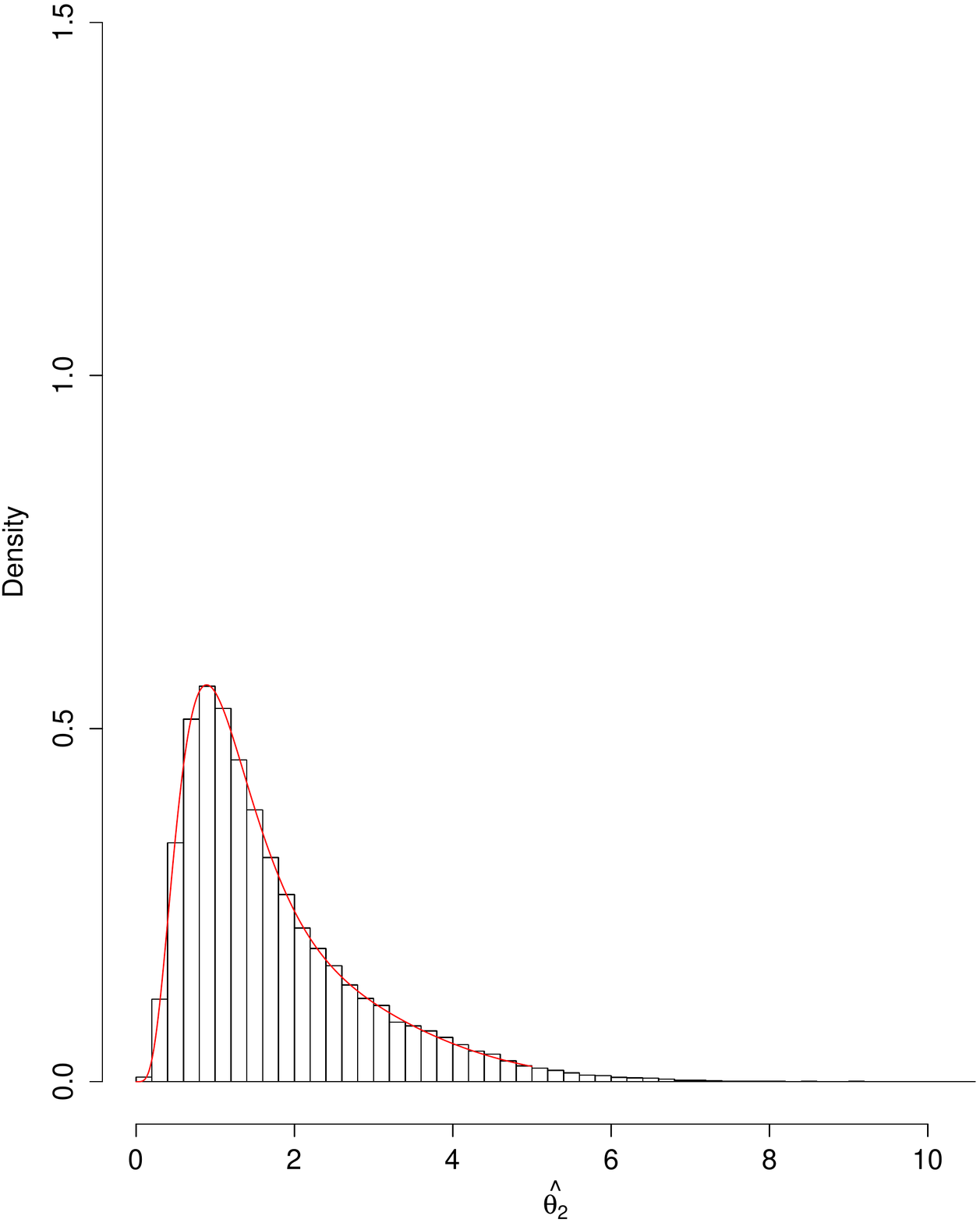}
  \caption{Histogram of $\what{\theta}_2$ along with its PDF}
  \label{fig:sfig2}
\end{subfigure}
\caption{Histogram of $\what{\theta}_1$ and $\what{\theta}_2$ along with its PDF,taking $\theta_1=.5$, $\theta_2=1.5$, $m=20$, $n=25$, $k=6$,$R=(2_{(5)})$}
\label{fig:fig4}
\end{figure}

Now to get an idea about the shape of the PDFs of $\widehat{\theta}_1$ and $\widehat{\theta}_2$, for different censoring
schemes, we have plotted in Figures \ref{fig:fig1} to \ref{fig:fig4} the PDFs of $\widehat{\theta}_1$ and $\widehat{\theta}_2$
along with the histograms of $\widehat{\theta}_1$ and $\widehat{\theta}_2$ based on 10,000 replications.  
%\enlargethispage{1 in}

\enlargethispage{1 in}

Some of the points are quite clear from the PDFs of $\widehat{\theta}_1$ and $\widehat{\theta}_2$.  The PDFs of both 
$\widehat{\theta}_1$ and $\widehat{\theta}_2$ are unimodal and are right 
skewed for different parameter values and for different sample sizes.  Moreover, in all the cases it is observed that the 
modes of the PDFs are very close to the corresponding true parameter values, as expected.

\section{\sc Generation of the Data and the Expected Experimental Time}

It is observed that for the proposed NJPC scheme, it is quite simple to generate samples for a given censoring scheme, hence 
simulation experiments can be performed quite efficiently.  In this section we provide an algorithm to generate sample from a
given NJPC scheme.  This algorithm is based on the following lemma.

\noindent {\sc Lemma 2:} If $W_1\leq \ldots \leq W_k$ are the ordered lifetime from a NJPC, then 
$$
W_i\overset{d}{=}\sum_{s=1}^{i}V_s,
$$
where $V_s$'s are independent random variables such that 
$$
V_s \sim \hbox{Exp} \left (\frac{1}{E_s} \right ), \ \ \
E_s=\frac{(m-\sum_{j=1}^{s-1}(R_j+1))}{\theta_1}+\frac{(n-\sum_{j=1}^{s-1}(R_j+1))}{\theta_2}.
$$
\noindent {\sc Proof:} See in the Appendix.   \qed

Now we can use the following algorithm to generate $({\bold W}, {\bold Z})$ for a given $n,m,k, R_1, \ldots, R_{k-1}$.

\noindent {\sc Algorithm:} 
\begin{itemize}
\item Step 1: Compute $E_s$, for $s=1,\ldots, k$.
\item Step 2: Generate $\ds V_s \sim \hbox{Exp} \left (\frac{1}{E_s} \right )$, $s=1,\ldots, k$.
\item Step 3: Compute $\ds W_i=\sum_{s=1}^{i}V_s, i=1,\ldots,k.$
\item Step 4: Generate $\ds Z_i\sim \hbox{Bin}(1,p_i), i = 1, \ldots, k$, where 
$$
p_i=\frac{(m-\sum_{j=1}^{i-1}(R_j+1))\theta_2}{(m-\sum_{j=1}^{i-1}(R_j+1))\theta_2+(n-\sum_{j=1}^{i-1}(R_j+1))\theta_1}.
$$
\end{itemize}
\qed

\noindent Using Lemma 2, we can easily obtain the expected experimental time as 
$$
E(W_k) = \sum_{s=1}^k E(V_s) = \sum_{s=1}^k \frac{1}{E_s}.
$$

%In simulation tables, to compute MSE, AE and Bootstrap confidence interval for our new censoring scheme, we apply the given algorithm every time.

\section{\sc Construction of Confidence Interval}
\subsection{\sc Exact Confidence Interval}

Based on the assumptions that $P_{\theta_1}(\widehat{\theta}_1>t)$ is a strictly increasing function of $\theta_1$ for any point 
$t > 0$ when $\theta_2$ is fixed, a $100(1-\alpha)\%$ exact confidence interval of $\theta_1$ can be constructed.  Similarly, 
based on the assumption that $P_{\theta_2}(\widehat{\theta}_2>t)$ is a strictly increasing function of $\theta_2$ 
for any point $t$ when $\theta_1$ is fixed,  a $100(1-\alpha)\%$ exact confidence interval of $\theta_2$ can be constructed as follows, see for example Lehmann and Romano \cite{LR:2005}.

Conditioning on $1\leq m_k\leq k-1$, a $100(1-\alpha)\%$ exact confidence interval for 
$\theta_1$ as ($\theta_{1L}, \theta_{1U}$) can be obtained by solving the following two nonlinear equations keeping 
$\theta_2$ fixed. 
\begin{eqnarray}
 \begin{cases}P_{\theta_{1L}}(\widehat{\theta}_1 > \widehat{\theta}_{1\textit {obs}}|1\leq m_k\leq k-1)=\frac{\alpha}{2},\\
P_{\theta_{1U}}(\widehat{\theta}_1 > \widehat{\theta}_{1\textit {obs}}|1\leq m_k\leq k-1)=1-\frac{\alpha}{2}.   \label{neq-1}
\end{cases} 
\end{eqnarray}
Similarly, conditioning on $1\leq m_k\leq k-1$, a $100(1-\alpha)\%$ exact confidence interval for ${\theta}_2$ as 
($\theta_{2L}, \theta_{2U}$) can be obtained by solving the following nonlinear equations keeping $\theta_1$ fixed. 
\begin{eqnarray}
 \begin{cases}P_{\theta_{2L}}(\widehat{\theta}_2 > \widehat{\theta}_{2\textit {obs}}|1\leq m_k\leq k-1)=\frac{\alpha}{2},\\
P_{\theta_{2U}}(\widehat{\theta}_2 > \widehat{\theta}_{2\textit {obs}}|1\leq m_k\leq k-1)=1-\frac{\alpha}{2}.   \label{neq-2}
\end{cases}
\end{eqnarray}
In practice to compute $(\theta_{1L}, \theta_{1U})$, we replace $\theta_2$ by its MLE $\widehat{\theta}_2$, similarly, 
to compute $(\theta_{2L}, \theta_{2U})$, we replace $\theta_1$ by its MLE $\widehat{\theta}_1$.  One can use the standard 
bisection method or Newton-Raphson method to solve these two \eqref{neq-1} and \eqref{neq-2} non-linear equations.

The following result provides the necessary monotonicity properties of $P_{\theta_1}(\widehat{\theta}_1>t)$ and 
$P_{\theta_2}(\widehat{\theta}_2>t)$.  It also justifies using \eqref{neq-1} and \eqref{neq-2} to construct the exact confidence
intervals of $\theta_1$ and $\theta_2$, respectively.

\noindent {\sc Lemma 3:}

\noindent (i) $P_{\theta_1}(\widehat{\theta}_1>t|1\leq m_k \leq k-1)$ is a strictly increasing function of 
$\theta_1$ for any point  $t$ when $\theta_2$ is kept fixed.

\noindent (ii) $P_{\theta_2}(\widehat{\theta}_2>t|1\leq m_k \leq k-1)$ is a strictly increasing function of 
$\theta_2$ for any point  $t$ when $\theta_1$ is kept fixed.

\noindent {\sc Proof:} See in appendix.   \qed

\subsection{\sc Bootstrap Confidence Interval}

Since the exact confidence intervals can be obtained by solving two non-linear equations we propose to use parametric 
bootstrap confidence intervals also as an alternative.  The following steps can be followed to construct parametric bootstrap
confidence intervals. 

\noindent Step 1: Given the original data, compute $\widehat{\theta}_1$, $\widehat{\theta}_2$.  \\
\noindent Step 2: Generate a bootstrap sample 
$\{({W_1}^{\ast},{Z_1}^{\ast}) \dots,({W_k}^{\ast},{Z_k}^{\ast})\}$ using the algorithm provided in Section 4 for 
a given $m$, $n$, $k$, $(R_1,\ldots R_{k-1})$, $\widehat{\theta}_1$, $\widehat{\theta}_2$,.  \\
\noindent Step 3: Compute $\widehat{{\theta}_1^{\ast}}$,  $\widehat{\theta}_2^{\ast}$ based on the bootstrap sample. \\
\noindent Step 4: Repeat Step 1-Step 3 say $B$ times and obtain $\{\widehat{\theta}_{11}^*, \ldots, \widehat{\theta}_{1B}^*\}$ 
and $\{\widehat{\theta}_{21}^*, \ldots, \widehat{\theta}_{2B}^*\}$.    Sort $\widehat{\theta}_{1j}^{\ast}$ in ascending order to get 
$\ds (\widehat{\theta}_{1(1)}^{\ast}, \ldots, \widehat{\theta}_{1(B)}^{\ast})$.
Similarly sort $\widehat{\theta}_{2j}^{\ast}$ in ascending order to get $\ds (\widehat{\theta}_{2(1)}^{\ast}, \ldots, 
\widehat{\theta}_{2(B)}^{\ast})$.  \\
\noindent Step 5: Construct a $100(1-\alpha)\%$ confidence interval for $\theta_1$
as $\big(\widehat{\theta}_{1([\frac{\alpha}{2}B])}^{\ast}, \widehat{\theta}_{1([(1-\frac{\alpha}{2})B])}^{\ast} \big)$ and a
$100(1-\alpha)\%$ confidence interval for $\theta_2$
as $\big(\widehat{\theta}_{2([\frac{\alpha}{2}B])}^{\ast}, \widehat{\theta}_{2([(1-\frac{\alpha}{2}B)])}^{\ast} \big)$.  Here $[x]$ 
denotes the largest integer less than or equal to $x$.

\section{\sc Simulation Results And Data Analysis}

\subsection{\sc Simulation Results}

We perform some simulation experiments to compare the performances of the estimators based on NJPC and JPC schemes.  We have taken
different $m$, $n$, $k$, different $(\theta_1, \theta_2)$  and different $R_1, \ldots, R_{k-1}$ values.  For a given set of parameters
and the sample sizes, we generate sample based on the algorithm provided in Section 4.  In each case we compute the 
MLEs based on the observed sample, and report their average estimates (AE) and mean squared 
errors (MSEs) based on 10,000 replications.  In each case for the NJPC scheme we construct the exact confidence intervals of $\theta_1$ and 
$\theta_2$, and we report the average lengths (AL) and the coverage percentages (CP) based on 1000 replications.  For each sample we compute the 
bootstrap confidence intervals based on 1000 replications and we report the average lengths and the coverage percentages based on 
1000 replications.  All the results are 
reported in Tables \ref{table-1} - \ref{table-4}.  We use the following notation to denote a particular progressive censoring 
scheme.  For example when $m$ = 15, $n$ = 12, $k$ = 6 and $R = (4, 0_{(4)})$ means $R_1$ = 4, $R_2 = R_3 = R_4 = R_5 = 0$.

\begin{table}[H]
\caption{ AE and MSE of the MLE's taking $\theta_1=.5, \theta_2=1, m=15,n=12$}\label{table-1}
\begin{center}
%\centering
\begin{tabular}{lllllll}
\toprule
\multirow{2}{*}{Censoring scheme} & \multirow{2}{*}{MLE} &\multicolumn{3}{l}{NJPC} & \multicolumn{2}{l}{JPC}\\ 
\cline{3-4}
\cline{6-7}
&&\multicolumn{1}{c}{AE} & \multicolumn{1}{c}{MSE} &\multicolumn{1}{c}{•} &\multicolumn{1}{c}{AE} &\multicolumn{1}{l}{MSE} \\
\midrule
k=6,R=(4,0$_{(4)}$) & $\what{\theta}_1$ & 0.575 & 0.099  &  &0.563 & 0.113\\
& $\what{\theta}_2$ & 0.995 & 0.377&  &1.125 & 0.607  \\
\midrule
k=6,R=(0,4,0$_{(3)}$) & $\what{\theta}_1$ & 0.577 & 0.106 &  &0.565 & 0.114  \\
& $\what{\theta}_2$ & 1.001 & 0.380&  &1.122 & 0.599 \\
\midrule
k=6,R=(0$_{(2)}$,4,0$_{(2)}$) & $\what{\theta}_1$ & 0.573 & 0.106 &  &0.571 & 0.112  \\
& $\hat{\theta}_2$ & 1.016 & 0.388 &  &1.147 & 0.622 \\
\midrule
k=6,R=(0$_{(3)}$,4,0) & $\hat{\theta}_1$ & .580 & 0.108 &  &0.567 & 0.112  \\
 & $\hat{\theta}_2$ & 1.034 & 0.411&  &1.133 & 0.598  \\
\midrule
k=6,R=(0$_{(4)}$,4) & $\hat{\theta}_1$ & 0.571 & 0.103 &  & 0.569 & 0.106  \\
& $\hat{\theta}_2$ & 1.044 & 0.421&  &1.124 & 0.585 \\
\bottomrule            
\end{tabular}
\end{center}
\end{table}
\enlargethispage{1 in}

\begin{table}[H]
\caption{ AE and MSE of the MLE's taking $\theta_1=.5$, $\theta_2=1$, $m=15$, $n=12$}\label{table-2}
\begin{center}
%\centering
\begin{tabular}{lllllll}
\toprule
\multirow{2}{*}{Censoring scheme} & \multirow{2}{*}{MLE} &\multicolumn{3}{l}{NJPC} & \multicolumn{2}{l}{JPC}\\ \cline{3-4}

\cline{6-7}

&&\multicolumn{1}{c}{AE} & \multicolumn{1}{c}{MSE} 
&\multicolumn{1}{c}{•}  &\multicolumn{1}{c}{AE} & \multicolumn{1}{l}{MSE} \\
\midrule
                        
k=8,R=(3,0$_{(6)}$) & $\what{\theta}_1$ & 0.538 & 0.056 &  & 0.537 & 0.062 \\

                        & $\what{\theta}_2$ & 1.121 & 0.504&  &1.238 & 0.838 \\
                        \midrule  
 
k=8,R=(0$_{(2)}$,3,0$_{(4)}$)  & $\what{\theta}_1$ & 0.541 & 0.059 &  & 0.534 & 0.063  \\
& $\what{\theta}_2$ & 1.134 & 0.523&  &1.226 & 0.805 \\
                        \midrule       
k=8,R=(0$_{(3)}$,3,$0_{(3)}$)     & $\what{\theta}_1$ & .539 & 0.056 &  & 0.534 & 0.061  \\
& $\what{\theta}_2$ & 1.138 & 0.543&  &1.238 & 0.817 \\
                        \midrule 
k=8,R=(0$_{(5)}$,7,0) & $\what{\theta}_1$ & 0.540 & 0.059 &  & 0.537 & 0.061  \\
 & $\what{\theta}_2$ & 1.156 & 0.577& &1.231 & 0.792\\
                        \midrule      
            k=8,R=(0$_{(6)}$,7)    & $\what{\theta}_1$ & 0.543 & 0.063 &  & 0.538 & 0.066  \\
& $\what{\theta}_2$ & 1.159 & 0.574&  &1.227 & 0.834 \\
                        \bottomrule
\end{tabular}
\end{center}
\end{table}

\begin{table}[H]
\caption{ AL and CP of CI's taking $\theta_1=.5, \theta_2=.6, m=20,n=25$}\label{table-3}
\begin{center}
%\centering
\begin{tabular}{lllllll}
\toprule
\multicolumn{1}{c}{Censoring scheme} & \multicolumn{1}{c}{Parameter}
&\multicolumn{3}{l}{Exact 90\% CI} & \multicolumn{2}{l}{Bootstrap 90\%CI}\\
\cline{3-4}

\cline{6-7}

&&\multicolumn{1}{c}{AL} & \multicolumn{1}{c}{CP}

&\multicolumn{1}{c}{}  &\multicolumn{1}{c}{AL} & \multicolumn{1}{l}{CP} \\

\midrule
k=8,R=(7,0$_{(6)}$) & $\theta_1$ & 2.920 &89.80\% & &1.279 &91.80\%    \\
& $\theta_2$ &2.190 &90.90\% & & 1.384 &89.00\%  \\
\midrule
k=8,R=(0$_{(3)}$,7,0$_{(3)}$) & $\theta_1$ & 2.912 &89.40\% & &1.288&90.70\%  \\
& $\theta_2$ &2.101 &91.70\% & &1.395&90.60\% \\
\midrule
k=8,R=(0$_{(5)}$,7,0)& $\theta_1$ &2.799& 88.80\% & &1.237&89.60\%    \\
& $\theta_2$ & 2.214 &91.40\%  &&1.479&91.10\%\\
                        \midrule
k=8,R=(0$_{(6)}$,7)& $\theta_1$ &2.871 &89.30\%  & & 1.246&89.50\%    \\
& $\theta_2$ &2.399 &90.50\%  &&1.409&89.20\%  \\
\midrule
 k=8,R=(0$_{(7)}$)& $\theta_1$ &2.476 &90.40\%  & &1.223&90.50\%    \\
& $\theta_2$ &2.455 &91.40\%  && 1.485 & 89.20\%  \\
                       
 \bottomrule

\end{tabular}
\end{center}
\end{table}

\begin{table}[H]
\caption{ AL and CP of CI's taking $\theta_1=.5, \theta_2=.6, m=20,n=25$} \label{table-4}
\begin{center}
%\centering
\begin{tabular}{lllllll}%\label{table-5}
\toprule
\multicolumn{1}{c}{Censoring scheme} & \multicolumn{1}{c}{Parameter} &\multicolumn{3}{l}{Exact 90\% CI} & \multicolumn{2}{l}{Bootstrap 90\%CI}\\
 \cline{3-4}

\cline{6-7}

&&\multicolumn{1}{c}{AL} & \multicolumn{1}{c}{CP}

&\multicolumn{1}{c}{•}  &\multicolumn{1}{c}{AL} & \multicolumn{1}{l}{CP} \\

\midrule
k=6,R=(10,0$_{(4)}$) & $\theta_1$ & 4.410 & 89.10\% &&1.213 &92.90\%     \\
& $\theta_2$ &3.188 &88.90\% & &1.531 &91.40\% \\
\midrule

 k=6,R=(0$_{(2)}$,10,0$_{(2)}$)& $\theta_1$ &4.252 &88.50\%& & 1.241&92.30\% \\
& $\theta_2$ & 3.201 &89.40\% & &1.578 & 90.80\% \\
  \midrule
 k=6,R=(0$_{(4)}$,10)& $\theta_1$ & 4.008 &88.40\% &&1.293 &91.70\%   \\
& $\theta_2$ &3.550& 90.90\%&& 1.543 &92.60\%  \\
                        \midrule
k=6,R=(0$_{(5)}$)& $\theta_1$ &3.642 &89.70\% & &1.253 &90.90\%   \\
& $\theta_2$ & 3.860 &90.10\% & & 1.511 &89.20\%  \\
\bottomrule            
 \end{tabular}
\end{center}
\end{table}

Some of the points are quite clear from the above Tables.  It is clear that for both the censoring schemes the estimators
are quite satisfactory.  In most of the cases considered here it is observed that the MSEs of both the estimators are smaller
in case of NJPC than the JPC.  Regarding the confidence intervals it is observed that the confidence intervals obtained
using the exact distribution and also using the bootstrap method provide satisfactory results.  In all the cases the coverage
percentages are very close to the nominal level.  Regarding the length of the confidence intervals, the bootstrap confidence
intervals perform slightly better than the exact confidence intervals.  Moreover, the implementation of the bootstrap method
is also quite simple in this case.  

Now we would like to discuss some of the computational issues we have encountered during the simulation experiments mainly
to calculate the exact confidence intervals of $\theta_1$ and $\theta_2$.  It is 
observed that for $m \ne n$, and when $k$ is large the computation of $P(X_r > t)$ and $P(Y_r > t)$ become quite difficult for
large value of $t$.  For small value of $k$, if $\theta_1$ and $\theta_2$ are quite different, then solving the two non-linear 
equations (\ref{neq-1}) and (\ref{neq-2}) become quite difficult.  
In this case 
$\ds P_{\theta_{1U}}(\widehat{\theta}_1 > \widehat{\theta}_{1\textit{obs}}|1 \le m_k \le k-1)$ and 
$\ds P_{\theta_{2U}}(\widehat{\theta}_2 > \widehat{\theta}_{2\textit{obs}}|1 \le m_k \le k-1)$  become very flat for large values of 
$\theta_{1U}$ and $\theta_{2U}$, respectively.  Hence the confidence intervals become very wide.  On the other hand the construction of confidence intervals based
on bootstrapping does not have any numerical issues.

Considering all these points we propose to use bootstrap method for constructing the 
confidence intervals in this case.

\subsection{\sc Data Analysis}

In this section we provide the analysis of a data set mainly for illustrative purposes.  These data sets were used by
Rasouli and Balakrishnan \cite{RB:2010} also and they were originally taken from Proschan \cite{Proschan:1963}.  The 
data represent the intervals between failures (in hours) of the air conditioning system of a fleet of 13 Boeing 720 jet
airplanes.  It is observed by  Proschan \cite{Proschan:1963} that the failure time distribution of the air conditioning system
for each of the planes can be well approximated by exponential distributions.  We have considered the planes ``7913'' and 
``7914'' for our illustrative purposes.  The data are presented below:

\noindent {\sc Plane 7914:} 3, 5, 5, 13, 14, 15, 22, 22, 23, 30,
        36, 39, 44, 46, 50, 72, 79, 88,
        97, 102, 139, 188, 197, 210.
  
\noindent {\sc Plane 7913:} 1, 4, 11, 16, 18, 18, 18, 24, 31, 39,
        46, 51, 54, 63, 68, 77, 80, 82,
        97, 106, 111, 141, 142, 163, 191,
        206, 216.

In this case $m$ = 24 and $n$ = 27.  We have considered two different NJPC with $k$ = 8, and different $R_i$ values.

\noindent {\sc Censoring Scheme 1:} $k=8$ and $R=(0_{(7)})$

Based on the above censoring scheme we generate ${\bf W}$ and ${\bf Z}$, and they are as follows. 
$w=(1, 3, 4, 5, 5, 11, 13, 15)$ $z=(0, 1, 0, 1, 1, 0, 1, 1)$.  We compute the MLEs of the unknown parameters and 
90\% exact and bootstrap confidence intervals in both the cases.  The results are reported in Table \ref{res-1}.

\begin{table}[h]
\caption{\sc Results related to Censoring Scheme 1.}   \label{res-1}
\begin{center}
%\centering
\begin{tabular}{lrll}
\toprule
\multicolumn{1}{c}{parameter} & \multicolumn{1}{c}{MLE} &
\multicolumn{1}{l}{Bootstrap 90\% CI} &\multicolumn{1}{l}{Exact 90\% CI}\\
\midrule
$\theta_1$&59.4 &(27.862,132.911)  & (30.027,141.049) \\
$\theta_2$&114.0 &(49.146,345.655) & (49.183,422.490)\\
\bottomrule
\end{tabular}
\end{center}
\end{table}

\noindent {\sc Censoring Scheme 2:} $k=8$ and $R=(2_{(7)})$

For the Censoring Scheme 2, the generated ${\bf W}$ and ${\bf Z}$ are $w=(1, 3, 4, 5, 5, 14, 15, 16)$ and 
 $z=(0, 1, 0, 1, 1, 1, 1, 0)$.  In this case the MLEs and the associate confidence intervals are reported in 
Table \ref{res-2}

\begin{table}[h]
\caption{Results related to Censoring Scheme 2.}   \label{res-2}
\begin{center}
%\centering
\begin{tabular}{lrll}
\toprule
\multicolumn{1}{c}{parameter} & \multicolumn{1}{c}{MLE} &
\multicolumn{1}{l}{Bootstrap 90\% CI} &\multicolumn{1}{l}{Exact 90\% CI}\\
\midrule
$\theta_1$&37.8 &(17.239,82.119)  & (19.318,93.453) \\
$\theta_2$&79.0 &(31.003,249.636) & (34.588,283.294)\\
\bottomrule
\end{tabular}
\end{center}
\end{table}

It is clear that the MLEs of the unknown parameters depend quite significantly on the censoring schemes, as expected.  The 
length of the confidence intervals based on bootstrapping are smaller than the exact confidence intervals.

\section{\sc Conclusion}

In this paper we introduce a new joint progressive censoring scheme for two samples.  Based on the assumptions that the 
lifetime distributions of the two populations follow exponential distributions we obtain the MLE's of the unknown parameters,
and derive their exact distributions.  It is observed that analytically the proposed model is easier to handle than the 
existing joint progressive censoring scheme of Rasouli and Balakrishnan \cite{RB:2010}.  We perform some simulation 
experiments and it is observed that in certain cases the MLEs of the unknown parameters based on the proposed model behave better
than the existing model.  Moreover, performing the simulation experiments based on the proposed model is easier compared to the 
existing model.  Therefore, the proposed model can be used for two sample problem quite conveniently in practice.

In this paper we have assumed that the lifetimes of the items follow exponential distribution.  In practice it may not 
be the case always because exponential distribution has a constant hazard rate.  It is well known that because of the 
flexibility, the Weibull distribution or the generalized exponential distribution are more useful in practice.  Therefore, 
it is important to develop the proper inferential procedures for other lifetime distributions for a two sample problem.  More
work is needed along these directions.

\section*{\sc Appendix}
\noindent{\sc Proof of Lemma 1:}  Note that 
\beanno 
P(m_k=r) & = & \sum_{\substack{\bold z}\in Q_r} P(Z_1=z_1,\ldots,Z_k=z_k)  \\
& = & \sum_{\substack{\bold z}\in Q_r} P(Z_1=z_1)P(Z_2=z_2|Z_1=z_1)\cdots P(Z_k=z_k|Z_{k-1}=z_{k-1},\ldots ,Z_1=z_1).
\eeanno
Now
\beanno
P(Z_i=z_i|Z_{i-1},\dots,Z_1=z_1) & = &\frac{(m-\sum_{j=1}^{i-1}(R_j+1))z_i+(n-\sum_{j=1}^{i-1}(R_j+1))(1-z_i)}{(m-\sum_{j=1}^{i-1}(R_j+1))p+(n-\sum_{j=1}^{i-1}(R_j+1))q} p^{z_i} q^{1-z_i},
\eeanno
where $\ds p=P(X<Y)=\frac{\theta_2}{\theta_1+\theta_2}$, $q=1-p$.  Hence $Z_i's$ are independent, therefore
\beanno
P(m_k=r)&=&\sum_{\substack{\bold z}\in Q_r} \prod_{i=1}^{k}\{\frac{(m-\sum_{j=1}^{i-1}(R_j+1))z_i+(n-\sum_{j=1}^{i-1}(R_j+1))(1-z_i)}{(m-\sum_{j=1}^{i-1}(R_j+1))p+(n-\sum_{j=1}^{i-1}(R_j+1))q} p^z_iq^{1-z_i}\}\\
& =& \sum_{\substack{\bold z}\in Q_r} {\prod_{i=1}^{k} \frac{(m-\sum_{j=1}^{i-1}(R_j+1))z_i+(n-\sum_{j=1}^{i-1}(R_j+1))(1-z_i)}{(m-\sum_{j=1}^{i-1}(R_j+1))p+(n-\sum_{j=1}^{i-1}(R_j+1))q} }p^{m_k} q^{n_k}\\
& =& \sum_ {\substack{\bold z}\in Q_r} {\prod_{i=1}^{k}\frac{(m-\sum_{j=1}^{i-1}(R_j+1))z_i+(n-\sum_{j=1}^{i-1}(R_j+1))(1-z_i)}{(m-\sum_{j=1}^{i-1}(R_j+1))\theta_2+(n-\sum_{j=1}^{i-1}(R_j+1))\theta_1}}{\theta_1}^{k-r}{\theta_2}^r.
\eeanno

\noindent{\sc Proof of Theorem 1:} Conditioning on $ 1\leq m_k \leq k-1$,
\beanno
M_{\what{\theta_1},\what{\theta_2}} \big(t_1,t_2\big) &=& E(e^{t_1\what{\theta_1} +t_2\what{\theta_2}}|1\leq m_k \leq k-1)\\
& = &\sum_{r=1}^{k-1}E(e^{t_1\what{\theta_1} + t_2\what{\theta_2}}|m_k=r)P(m_k=r|1 \leq m_k \leq k-1)\\
& = & \sum_{r=1}^{k-1}\sum_{\substack{\bold z}\in Q_r}
 E(e^{t_1\what{\theta_1} +t_2\what{\theta_2}}|m_k=r,\bold Z=\bold z)P(\bold Z=\bold z|m_k=r)P(m_k=r|1\leq m_k \leq k-1)\\
& = & \frac{C}{P(1 \leq m_k \leq k-1)}  \sum_{r=1}^{k-1}\sum_{\substack{\bold z}\in Q_r} C\frac{1}{{\theta_1}^r}\frac{1}{{\theta_2}^{k-r}}
\times   \\
&  &
 \int\limits_0^\infty\ \int\limits_{w_1}^{\infty}\ldots \int\limits_{w_{k-1}}^{\infty}   
e^{\frac{t_1 \{\sum_{i=1}^{k-1}(R_i+1)w_i+(m-\sum_{i=1}^{k-1}(R_i+1))w_k \}}{r}} \times e^{\frac{ t_2\{\sum_{i=1}^{k-1}(R_i+1)w_i+(n-\sum_{i=1}^{k-1}(R_i+1))w_k\}}{k-r}} \\
&  &  
\times e^{-\frac{1}{\theta_1}\{\sum_{i=1}^{k-1}(R_i+1)w_i+(m-\sum_{i=1}^{k-1}(R_i+1))w_k\}}\\
&  & \times e^{-\frac{1}{\theta_2}\{\sum_{i=1}^{k-1}(R_i+1)w_i+(n-\sum_{i=1}^{k-1}(R_i+1))w_k\}}
 \hspace{0.2mm}dw_k \ldots\hspace{0.2mm}dw_2\hspace{0.2mm} dw_1\\
& = &\frac{1}{P(1 \leq m_k \leq k-1)} \sum_{r=1}^{k-1}\sum_{\substack{\bold z}\in Q_r} C\frac{1}{{\theta_1}^r}\frac{1}{{\theta_2}^{k-r}}\\
& & { \{ \prod_{j=1}^{k}{\frac{(m-\sum_{i=1}^{j-1}(R_i+1))}{\theta_1} +\frac{(n-\sum_{i=1}^{j-1}(R_i+1))}{\theta_2}} \} }^{-1} \times\prod_{s=1}^{k}{(1-\alpha_{sr} t_1 -\beta_{sr}t_2)}^{-1} \\
 & = &\frac{1}{P(1 \leq m_k \leq k-1)}\sum_{r=1}^{k-1}P(m_k=r) \prod_{s=1}^{k}{(1-\alpha_{sr} t_1 -\beta_{sr}t_2)}^{-1}.
\eeanno 

\noindent{\sc Proof of Lemma 2:}
\beanno
E(e^{tW_j})& =& \sum_{r=0}^{k}\sum_{\substack{\bold z}\in Q_r}E(e^{tW_j}|m_k=r,\bold Z=\bold z)P(\bold Z=\bold z|m_k=r)P(m_k=r)\\
& = & C \sum_{r=1}^{k-1}\sum_{\substack{\bold z}\in Q_r}\int\limits_0^\infty\ \int\limits_{w_1}^{\infty}\ldots \int\limits_{w_{k-1}}^{\infty}\frac{1}{{\theta_1}^r}\frac{1}{{\theta_2}^{k-r}} e^{tw_j} \times e^{-\frac{1}{\theta_1} \{ \sum_{i=1}^{k-1}(R_i+1)w_i+(m-\sum_{i=1}^{k-1}(R_i+1))w_k \} }  \\
& &  \times e^{-\frac{1}{\theta_2} \{\sum_{i=1}^{k-1}(R_i+1)w_i+(n-\sum_{i=1}^{k-1}(R_i+1))w_k \} } \hspace{0.2mm}dw_k \ldots\hspace{0.2mm}dw_2\hspace{0.2mm} dw_1\\
& = & C \sum_{r=0}^{k}\sum_{\substack{\bold z}\in Q_r}\frac{1}{{\theta_1}^r}\frac{1}{{\theta_2}^{k-r}}\\
& & \times { \{ a_k(a_k+a_{k-1})\cdots(a_k+a_{k-1}+\cdots +a_{j+1})
(a_k+a_{k-1}+\cdots + a_{j+1}+a'_j+a_{j-1})\}}^{-1}\cdots \\
& & \times {(a_k+a_{k-1}+\cdots +a_{j+1}+a'_j+a_{j-1}+\cdots +a_1) }^{-1} \\
& =& C \sum_{r=0}^{k}\sum_{\substack{\bold z}\in Q_r}\frac{1}{{\theta_1}^r}\frac{1}{{\theta_2}^{k-r}}\\
& & \times { \{ a_k(a_k+a_{k-1})\cdots(a_k+a_{k-1}+\cdots + a_j)\cdots(a_k+a_{k-1}+\cdots +a_j+a_{j-1}+\cdots a_1) \} }^{-1} \\
& & \times \frac{(a_k+a_{k-1}+\cdots +a_j)\cdots (a_k+a_{k-1}+\cdots +a_j+a_{j-1}+\cdots +a_1)}{(a_k+a_{k-1}+\cdots +a'_j)\cdots (a_k+a_{k-1}+\cdots +a'_j+a{j-1}+\cdots +a_1)}\\
& =& \sum_{r=0}^{k}P(m_k=r) \left \{ \frac{(a_k+a_{k-1}+\cdots +a_j)\cdots (a_k+a_{k-1}+\cdots +a_j+a_{j-1}+\cdots +a_1)}{(a_k+a_{k-1}+\cdots +a'_j)\cdots (a_k+a_{k-1}+\cdots +a'_j+a_{j-1}+\cdots +a_1)} \right \} \\
& = &  \left \{ \frac{(a_k+a_{k-1}+\cdots +a_j)\cdots (a_k+a_{k-1}+\cdots +a_j+a_{j-1}+\cdots +a_1)}{(a_k+a_{k-1}+\cdots +a'_j)\cdots (a_k+a_{k-1}+\cdots +a'_j+a_{j-1}+\cdots +a_1)} \right \} \sum_{r=0}^{k}P(m_k=r) \\
& =& \prod_{s=1}^j \left (1-\frac{t}{E_s} \right )^{-1}.  
\eeanno
Here
\beanno 
a_j & = & \frac{(R_j+1)}{\theta_1} + \frac{(R_j+1)}{\theta_2}, \ \ \ j=1,\ldots,k-1;  \\
a_k & = & \frac{(m-\sum_{j=1}^{k-1}(R_j+1))}{\theta_1} + \frac{(n-\sum_{j=1}^{k-1}(R_j+1))}{\theta_2}; \ \ \ a'_j=a_j-t;  \\
E_s & =  & \frac{(m-\sum_{j=1}^{s-1}(R_j+1))}{\theta_1}+\frac{(n-\sum_{j=1}^{s-1}(R_j+1))}{\theta_2}.
\eeanno

\noindent{\sc Proof of Lemma 3:} To prove Lemma 3, we mainly use the ``Three Monotonicity Lemmas'' of 
Balakrishnan and Iliopoulos \cite{BI:2009}.  We briefly state the ``Three Monotonicity Lemmas'' for convenience, and 
we will show that both $\widehat{\theta}_1$ and $\widehat{\theta}_2$ satisfy the ``Three Monotonicity Lemmas''.

Suppose $\widehat{\theta}$ is an estimate of $\theta$, and the survival function of $\widehat{\theta}$ can be written in the
following form: 
$$
P_{\theta}(\what{\theta}>x)=\sum_{\substack{d}\in{ {\cal D}}}P_{\theta}(\what{\theta}>x|D=d)P_{\theta}(D=d),
$$
where ${\cal D}$  is a finite set.

\noindent {\sc Lemma} (Three Monotonicity Lemmas:)  Assume that the following hold true: \\
\noindent (M1) $P_{\theta}(\what{\theta}>x|D=d)$ is increasing in $\theta$ for all $x$ and $d \in {\cal D}$;  \\
\noindent (M2) For all $x$ and $\theta>0$, $P_{\theta}(\what{\theta}>x|D=d)$ is decreasing in $d \in {\cal D}$;  \\
\noindent (M3) $D$ is stochastically decreasing in $\theta$.  \\
\noindent Then $P_{\theta}(\widehat{\theta}>x)$ is increasing in $\theta$ for any fixed $x$.

\noindent Now to prove (i), first observe that 
$$
P_{\theta_1}(\what{\theta}_1>t|1\leq m_k\leq k-1)=\sum_{r=1}^{k-1}P_{\theta_1}(\what{\theta}_1 >t|m_k=r)P_{\theta_1}(m_k=r|1\leq m_k \leq k-1).
$$
Hence, (i) can be proved if we can show that \\
\noindent (M1) $P_{\theta_1}(\what{\theta}_1>t|m_k=r)$ is increasing in $\theta_1$, $\forall t,r\in \{1,\ldots k-1\}$;  \\
\noindent (M2) $P_{\theta_1}(\what{\theta}_1>t|m_k=r)$ is  decreasing in $r$, $\forall t,\theta_1>0$;  \\
\noindent (M3) The conditional distribution of $m_k$ is stochastically decreasing in $\theta_1$.

\noindent From the  moment generating function of $\ds E(e^{t\widehat{\theta}_1}|m_k=r)$ it is easily observe that conditioning on 
$m_k=r$, $\widehat{\theta}_1\overset{d}{=}\sum_{s=1}^{k}X_{sr}$, where $X_{sr} \sim \hbox{Exp}(\alpha_{sr})$ and they are 
independently distributed.  Here $\alpha_{sr}$'s are same as defined in Theorem 1.  Since $\alpha_{sr}$ is increasing with $\theta_1$,
the distribution of $X_{sr}$ is stochastically increasing with $\theta_1$.  Since $X_{sr}$'s are independently distributed,  
(M1) is satisfied.

Now to prove (M2), observe that 
\beanno
\widehat{\theta_1}|\{m_k=r\} &   \overset{d}{=} & \frac{\sum_{i=1}^{k-1}{(R_i+1)w_i} +(m-\sum_{i=1}^{k-1}
(R_i+1))w_k}{r}  \\
\widehat{\theta_1}|\{m_k=r+1\}  &  \overset{d}{=} & \frac{\sum_{i=1}^{k-1}{(R_i+1)w_i} +(m-\sum_{i=1}^{k-1}
(R_i+1))w_k}{r+1}.
\eeanno
Hence for all $t$ and for $\theta_1 > 0$, 
$\ds P_{\theta_1}(\what{\theta}_1>t|m_k=r)>P_{\theta_1}(\what{\theta}_1>t|m_k=r+1)$.  This proves (M2).
   
To prove (M3) it is enough to show $m_k$ has monotone likelihood ratio property with respect to $\theta_1$.  For 
$\theta_1<{\theta_1}'$
$$
\frac{P_{\theta_1}(m_k=r|1\leq m_k \leq k-1)}{P_{{\theta_1}'}(m_k=r|1\leq m_k \leq k-1)} \propto \frac{P_{\theta_1}(m_k=r)}{P_{{\theta_1}'}(m_k=r)}
\propto {\big(\frac{\theta_1}{{\theta_1}'}\big)}^{k-r} \uparrow r.
$$

%\clearpage

\end{document}